# LOPSIDED ELASTIC DUMBBELL SUSPENSION

Nhan Phan-Thien (Phan Thiện Nhân)[1,2], M.A. Kanso (منى قانصو)[3,*] and A.J. Giacomin[4,5]

[1]State Key Laboratory of Fluid Power and Mechatronic Systems and Department of Engineering Mechanics, Zhejiang University, Hangzhou 310027, China

[2]Department of Mechanical Engineering, National University of Singapore, Singapore 117575, Singapore

[3]Chemical Engineering Department, Massachusetts Institute of Technology, Cambridge 02139, USA

[4]Mechanical Engineering Department, University of Nevada, Reno, Nevada 89557-0312, USA

[5]State Key Laboratory for Turbulence and Complex Systems, Peking University, Beijing 100871, China

*Corresponding author (kansom@mit.edu)

## ABSTRACT

We derive the constitutive equation for a suspension of lopsided Hookean dumbbells. By *lopsided*, we mean that one bead is larger than the other. We find that all results derived for symmetric Hookean dumbbells can be taken over for lopsided ones by replacing $2/\zeta$ with $(1/\zeta_1) + (1/\zeta_2)$, where $\zeta$, $\zeta_1$ and $\zeta_2$ are the bead friction coefficients for the symmetric dumbbell beads, and the first and second beads of the lopsided dumbbell respectively.

---

Much of the behaviors of elastic liquids observed in the laboratory can be explained by considering the orientation distributions of the macromolecules in suspension [1]. In macromolecular theory, the simplest relevant model of macromolecules is the dumbbell suspension. By *dumbbell*, we mean an object consisting of two and only two beads. If the bead separation is constant, the dumbbell is called *rigid*. The theory of rigid dumbbells explains rheological behaviors by virtue of macromolecular orientation alone ([1]; **Chapter 14** of [2]). If the bead separation is proportional to the component of force along the macromolecule, the dumbbell is called *Hookean* (Subsection 13.4 of [2]).

Figure 1 illustrates a lopsided Hookean dumbbell. The Hookean dumbbell thus offers the prospect of exploring the role of macromolecular extensibility on the rheology of a polymeric liquid. While Hookean dumbbells can lengthen and shorten, they can neither twist nor bend. Table 1 gives nomenclature, symbols and dimensions for our variables, both dimensional and dimensionless.

If the dumbbell beads are identical, the dumbbell is called *symmetric*, and otherwise, *lopsided*. Driven by curiosity, Abdel-Khalik and Bird [3] explored dumbbell lopsidedness, and for rigid dumbbells, discovered that all results for symmetric rigid dumbbells can be taken over for lopsided ones with the replacement (Rows 2 of **TABLE 13.4-1** of [4] or **TABLE 16.4-1** of [2]):

$$\zeta = \frac{2\zeta_1\zeta_2}{\zeta_1 + \zeta_2} \quad (1)$$

or:

$$\frac{2}{\zeta} = \frac{1}{\zeta_1} + \frac{1}{\zeta_2} \quad (2)$$

For instance, the relaxation time of a symmetric rigid dumbbell (Eq. (6) of [5]; [6]):

$$\lambda_R = \frac{\zeta L^2}{12kT} \quad (3)$$

for its lopsided rigid counterpart, becomes:

$$\lambda_R^\ell = \frac{\zeta_1\zeta_2}{\zeta_1 + \zeta_2}\left(\frac{L^2}{6kT}\right) \quad (4)$$

and the polymer contribution to the zero-shear viscosity (Eq. (6.7) of [1]):

$$\eta_0 - \eta_s = nkT\lambda_R \quad (5)$$

becomes:

$$\eta_0 - \eta_s = nkT\frac{\zeta_1\zeta_2}{\zeta_1 + \zeta_2}\left(\frac{L^2}{6kT}\right) \quad (6)$$

Eqs. (2) and (3) imply (Eq. (103) of [7]):

$$\frac{2}{a} = \frac{1}{a_1} + \frac{1}{a_2} \quad (7)$$

For a comparison of rheological material functions for a suspension of symmetric rigid dumbbells *versus* symmetric Hookean dumbbells, see § 24. of [1] and compare rows 1 and 3 of **TABLE 16.4-1** of [2]. We devote this Letter to exploring the effect of lopsidedness on a suspension of Hookean dumbbells, for which the expression for the relaxation time for the symmetric special case (Eq. (7.51) of [8] (or Eqs. (7.57) of [9,10]):

$$\lambda_H = \frac{\zeta}{4H} \quad (8)$$

is well-known. In the limit as $H \to \infty$, Eq. (8) does not recover Eq. (3). Otherwise put, the symmetric Hookean dumbbell [**Error! Reference source not found.** (lower right)] is not an extremum of its symmetric rigid counterpart [**Error! Reference source not found.** (upper left)].

We consider a simple elastic lopsided dumbbell model, where all the interactions between the suspending solvent and the dumbbell beads act at the bead centers, at position vectors $\mathbf{r}_1$ and $\mathbf{r}_2$. Each bead $i = 1,2$ is associated with a radius $a_i$, a mass $m_i$ and with a Stokes frictional coefficient:

$$\zeta_i = 6\pi\eta_s a_i \quad (9)$$

where $\eta_s$ is the solvent viscosity. Let $i = 2$ designate the second of the two beads, and let the *dumbbell lopsidedness* be the ratio:

$$\nu \equiv \frac{a_2}{a_1} \quad (10)$$

and thus, $\zeta_2 = \nu\zeta_1$. So, if $a_2 > a_1$, then $\nu > 1$ also. Results for the symmetric dumbbell must therefore be recovered when $\nu = 1$.

We connect the two beads with a massless dimensionless linear spring of constant stiffness, $H$. The beads, each subject to Newton's second law of motion, are acted on by the (i) fluid viscous resistance force, $\zeta_i(\mathbf{u}_i - \dot{\mathbf{r}}_i)$ with $\mathbf{u}_i = \mathbf{u}(\mathbf{r}_i)$ being the fluid velocity evaluated at bead $i$, (ii) elastic restoring force between the two beads, $\pm H(\mathbf{r}_2 - \mathbf{r}_1)$ and (iii) random force $\mathbf{F}_i^{(b)}(t)$ of Brownian motion, so that for $i = 1$ and $i = 2$ respectively (Eq. (7.40) of [8]; Eqs. (7.46) of [9,10]):

$$m_1 \ddot{\mathbf{r}}_1 = \zeta_1(\mathbf{u}_1 - \dot{\mathbf{r}}_1) + H(\mathbf{r}_2 - \mathbf{r}_1) + \mathbf{F}_1^{(b)}(t)$$
$$m_2 \ddot{\mathbf{r}}_2 = \zeta_2(\mathbf{u}_2 - \dot{\mathbf{r}}_2) + H(\mathbf{r}_1 - \mathbf{r}_2) + \mathbf{F}_2^{(b)}(t)$$
(11)

We call this dynamical system the *Langevin equations* (Eq. (3) and (4) of [11], [12]). Both Brownian forces have zero means, uncorrelated to each other, and their strength (or autocorrelation) is given by a fluctuation-dissipation theorem which relates this strength to the bead mobility (or bead friction coefficient) (Eq. (7.46) of [8]; Eqs. (7.47) of [9,10]):

$$\langle \mathbf{F}_j^{(b)}(t) \rangle = \mathbf{0}, \langle \mathbf{F}_1^{(b)}(t+s)\mathbf{F}_1^{(b)}(t) \rangle = 2kT\zeta_1 \delta(s)\mathbf{I}$$
$$\langle \mathbf{F}_1^{(b)}(t+s)\mathbf{F}_2^{(b)}(t) \rangle = \mathbf{0}, \langle \mathbf{F}_2^{(b)}(t+s)\mathbf{F}_2^{(b)}(t) \rangle = 2kT\zeta_2 \delta(s)\mathbf{I}$$
(12)

where $\delta(s)$ is the *Dirac delta function*, $s$ is an interval of past time, $\mathbf{I}$ is the identity tensor, and in which the chevrons, $\langle \ \rangle$, denote ensemble averages of their enclosures.

The dumbbell center of resistance, $\mathbf{R}^{(r)}$, and the dumbbell center-to-center vector $\mathbf{R}$ are related by:

$$(\zeta_1 + \zeta_2)\mathbf{R}^{(r)} = \zeta_1 \mathbf{r}_1 + \zeta_2 \mathbf{r}_2, \mathbf{R} = \mathbf{r}_2 - \mathbf{r}_1 \tag{13}$$

Since $\zeta_2 = \nu \zeta_1$, the center of resistance is also $(1 + \nu)\mathbf{R}^{(r)} = \mathbf{r}_1 + \nu \mathbf{r}_2$. By *center of resistance*, we mean the point on the dumbbell where a single applied force results in pure translation. Also, letting $\mathbf{R}^{(i)}$ be the vector from the center of resistance to the center of bead $i$, we get:

$$\mathbf{r}_1 = \mathbf{R}^{(r)} - \mathbf{R}^{(1)}, \mathbf{r}_2 = \mathbf{R}^{(1)} - \mathbf{R}^{(2)}, \mathbf{R} = \mathbf{R}^{(1)} - \mathbf{R}^{(2)} \tag{14}$$

and, by the definition of the center of resistance:

$$\zeta_2 \mathbf{R}_2 - \zeta_1 \mathbf{R}_1 = \zeta_2 \mathbf{r}_2 - \zeta_2 \mathbf{R}^{(r)} + \zeta_1 \mathbf{r}_1 - \zeta_2 \mathbf{R}^{(r)} = \zeta_1 \mathbf{r}_1 + \zeta_2 \mathbf{r}_2 - (\zeta_1 + \zeta_2)\mathbf{R}^{(r)} = 0 \tag{15}$$

There are three-time scales in the system: (i) the relaxation time scale $O(\zeta_i H^{-1})$, (ii) the inertial time scale $O(m\zeta_i^{-1})$, and (iii) the correlation time scale of the Brownian forces. Usually, the relaxation time scale dwarfs the inertial $[O(\zeta_i H^{-1}) \gg O(m_i \zeta_i^{-1})]$, so we customarily neglect bead inertial terms in Eq. (12). The equations of motion for the center of resistance, and the bead center-to-center vector then become:

$$(\zeta_1 + \zeta_2)\dot{\mathbf{R}}^{(r)} = \zeta_1 \mathbf{u}_1 + \zeta_2 \mathbf{u}_2 + \mathbf{F}^{(b,cr)}(t)$$
$$\zeta_2 \dot{\mathbf{R}} = \zeta_2(\mathbf{u}_2 - \mathbf{u}_1) - H(1 + \nu)\mathbf{R} + \mathbf{F}^{(b,R)}(t)$$
(16)

where $\mathbf{F}^{(b,cr)}(t)$ is the Brownian force acting on the center of resistance, and $\mathbf{F}^{(b,R)}(t)$ is the Brownian force acting on the center-to-center vector. These are given by:

$$\mathbf{F}^{(b,cr)}(t) = \mathbf{F}_1^{(b)}(t) + \mathbf{F}_2^{(b)}(t) \tag{17}$$

$$\mathbf{F}^{(b,R)}(t) = \mathbf{F}_2^{(b)}(t) - \nu \mathbf{F}_1^{(b)}(t) \tag{18}$$

which, along with Eq. (12), imply the statistics:

$$\langle \mathbf{F}^{(b,cr)}(t) \rangle = 0, \langle \mathbf{F}^{(b,cr)}(t+s)\mathbf{F}^{(b,cr)}(t) \rangle = 2kT(\zeta_1 + \zeta_2)\delta(s)\mathbf{I}$$
$$\langle \mathbf{F}^{(b,R)}(t) \rangle = 0, \langle \mathbf{F}^{(b,R)}(t+s)\mathbf{F}^{(b,R)}(t) \rangle = 2kT(\zeta_2 + \nu^2 \zeta_1)\delta(s)\mathbf{I}$$
(19)

Expanding the solvent velocity in a Taylor series about the center of resistance:

$$\begin{aligned}\mathbf{u}_1 &= \mathbf{u}^{(r)} - \mathbf{R}^{(1)} \cdot \nabla \mathbf{u}^{(r)} + \mathbf{R}^{(1)}\mathbf{R}^{(1)} : \nabla\nabla \mathbf{u}^{(r)} + HOT \\ \mathbf{u}_2 &= \mathbf{u}^{(r)} - \mathbf{R}^{(2)} \cdot \nabla \mathbf{u}^{(r)} + \mathbf{R}^{(2)}\mathbf{R}^{(2)} : \nabla\nabla \mathbf{u}^{(r)} + HOT\end{aligned} \quad (20)$$

in which the superscripted "$(r)$" denotes an evaluation at the center of resistance, $\nabla \mathbf{u}^{(r)}$ is the velocity gradient evaluated at the dumbbell center of resistance, and $HOT$ are the higher order terms. Using Eq. (14) and neglecting quadratic and higher-order terms, the Langevin equations, Eqs. (12), then finally become (Eq. (7.47) of [8]; Eqs. (7.53) of [9,10]):

$$\begin{aligned}\dot{\mathbf{R}}^{(r)} &= \mathbf{u}^{(r)} + (\zeta_1 + \zeta_2)^{-1} \mathbf{F}^{(b,cr)}(t) \\ \dot{\mathbf{R}} &= \mathbf{L} \cdot \mathbf{R} - (1+\nu)\zeta_2^{-1} H \mathbf{R} + \zeta_2^{-1} \mathbf{F}^{(b,R)}(t)\end{aligned} \quad (21)$$

where $\mathbf{L} \equiv \left(\nabla \mathbf{u}^{(r)}\right)^T$ is the transpose of the velocity gradient evaluated at the center of resistance, and the statistics of the random forces are given in Eqs. (19).

If the flow is homogeneous, $\mathbf{L}$ is a tensor-valued constant, $\nabla\nabla \mathbf{u}^{(r)}$ and the $HOT$ are identically zero, and the dumbbell center of resistance drifts just like a fluid particle on the mean:

$$\langle \dot{\mathbf{R}}^{(r)}\rangle = \langle \mathbf{u}^{(r)}\rangle = \mathbf{L}\langle \mathbf{R}^{(r)}\rangle = \mathbf{u}(\langle \mathbf{R}^{(r)}\rangle) \quad (22)$$

Migration from the streamline of the center of gravity is induced by a heterogeneous flow field. This migration is proportional to $\nabla\nabla \mathbf{u}^{(r)}$, neglecting the $HOT$.

## I. RESULTS

In this Letter, we proceed pedagogically with a mean motion analysis. By *mean motion*, we mean using mean center-to-center vectors for the bead positions. However, the diffusion equation underpinning the Langevin equations [Eqs. (21)], Eqs. (7.29) of [9,10] (or Eq. (7.28) of [8]), has been derived (see § 7.3.5 of [8,9,10]), and thus our findings will apply exactly for any statistic, not just the mean motion that we will use presently. The mean center-to-center vector thus evolves in time according to:

$$\langle \dot{\mathbf{R}}\rangle = \mathbf{L}\langle \mathbf{R}\rangle - (1+\nu)\zeta_2^{-1} H \langle \mathbf{R}\rangle = L\langle \mathbf{R}\rangle - \left(2\lambda_H^\ell\right)^{-1}\langle \mathbf{R}\rangle \quad (23)$$

which consists of flow-induced stretching, $\mathbf{L}\langle \mathbf{R}\rangle$, *minus* a restoring mechanism due to the connector spring force [$(1+\nu)\zeta_2^{-1}H\langle \mathbf{R}\rangle$ or $\left(2\lambda_H^\ell\right)^{-1}\langle \mathbf{R}\rangle$] and in which $\lambda_H^\ell$ is the relaxation time of the lopsided Hookean dumbbell:

$$\lambda_H^\ell \equiv \frac{2\zeta_2}{4(1+\nu)H} = \frac{2\nu\zeta_2}{4(1+\nu)H} = \frac{\zeta_1\zeta_2}{2(\zeta_1+\zeta_2)H} \quad (24)$$

This relaxation time accounts for the evolutions of both the orientation and the conformation of the lopsided dumbbell. It reduces to the well-known expression for the relaxation time of the symmetric Hookean dumbbell, Eq. (8), as it must. Evaluating this for the second bead, and then using the result to non-dimensionalize Eq. (24), we get:

$$\frac{\lambda_H^\ell}{\lambda_H} \equiv \frac{2}{1+\nu} \quad (25)$$

from which we learn that, when compared to its symmetric counterpart, referenced to its second bead, $\lambda_H = \zeta_2/4H$, the lopsided dumbbell relaxation time differs from the symmetric by the factor $2/(1+\nu)$. Figure 2 illustrates this. Eq. (24) [or its dimensionless companion Eq. (25)] are main results of this Letter.

The effect of lopsidedness of a Hookean dumbbell is thus arrived at immediately by replacing $\zeta$ in any expression derived for its symmetric counterpart with:

$$\zeta = \frac{2\zeta_1\zeta_2}{\zeta_1 + \zeta_2} \tag{26}$$

or:

$$\frac{2}{\zeta} = \frac{1}{\zeta_1} + \frac{1}{\zeta_2} \tag{27}$$

which match, surprisingly, the well-known results for the lopsided rigid dumbbells of Eqs. (1) and (2). Eq. (26) is a main result of this Letter.

For instance, the well-known result for the relaxation time of a symmetric Hookean dumbbell suspension [Eq. (8)], for lopsided Hookean dumbbells, becomes Eq. (24), and the polymer contribution to the zero-shear viscosity (Eqs. (7.56) and (7.60) of [8]; Eqs. (7.63) and (7.67) of [9,10]):

$$\eta_0 - \eta_s = nkT\lambda_H \tag{28}$$

becomes:

$$\eta_0 - \eta_s = nkT\lambda_H^\ell = \frac{nkT}{2H}\frac{\zeta_1\zeta_2}{\zeta_1 + \zeta_2} \tag{29}$$

for lopsided Hookean dumbbells.

Alternatively, since the bead friction coefficients are proportional to the bead radii, we can take over results for symmetric Hookean dumbbells by replacing the radius of the symmetric dumbbell using:

$$\frac{2}{a} = \frac{1}{a_1} + \frac{1}{a_2} \tag{30}$$

which, again surprisingly, matches the well-known result for lopsided rigid dumbbells of Eq. (7). Further, the constitutive equation for symmetric Hookean dumbbells (Eq. (7.55) of [8]; Eqs. (7.62) of [9,10]; Eq. (31) [with Eq. (32) of [13]; [14]; Eqs. (4.23) or (4.24) of [1]; Eq. (5.53) of [15]; Eq. (13.4-4) of [4]) can also be taken over yielding:

$$\mathbf{S}^{(p)} + \lambda_H^\ell \left\{ \frac{d}{dt}\mathbf{S}^{(p)} - \mathbf{L}\mathbf{S}^{(p)} - \mathbf{S}^{(p)}\mathbf{L}^T \right\} = nkT\,\mathbf{I} \tag{31}$$

in which $d/dt$ is the substantial derivative, $\mathbf{S}^{(p)}$ is the contribution of lopsided Hookean dumbbells to the total stress tensor, $\mathbf{S}^{(s)}$ is the solvent contribution, and in which the braces, { }, enclose the upper convected derivative of $\mathbf{S}^{(p)}$. In the symbols of the Oldroyd 8-constant framework (Eq. (8.1-2) of [16]; Eq. (3) of [17]), for the lopsided Hookean dumbbell, we recover Eq. (31) with (i) Eq. (29) for $\eta_0$, (ii) $\lambda_1 = \mu_1 = \lambda_H^\ell$, (iii) $\lambda_2 = \mu_2 = [\eta_s/(\eta_s + nkT\lambda_H^\ell)]\lambda_H^\ell$ and (iv) $\mu_0 = \nu_1 = \nu_2 = 0$ (see Eqs. (10.4-7) through (10.4-9) of [18]; **TABLE 8.1-1** of [16]; **Table 1** of [17]). We call this special case the *Oldroyd-B model* (see Subsections **7.5.5** of [8, 9,10]). The equivalent integral form of the constitutive equation for symmetric Hookean dumbbells (Eq. (5.58) of [15]; Eq. (13.4-9) of [4]), can also be taken over for lopsided Hookean dumbbells. From Eq. (31) we learn that, for lopsided Hookean dumbbell suspensions, the *elastic shear modulus*, $nkT$, is unaffected by dumbbell lopsidedness.

We have learnt that the results for lopsided Hookean dumbbell suspensions follow the same replacement relations as rigid dumbbell suspensions. By this we mean that any result for a symmetric Hookean dumbbell suspension can be taken over by use of the replacement relation for the bead friction coefficient, Eq. (26) [or alternatively, its implication for the bead radius, Eq. (30)]. Eq. (26) is thus a main result of this Letter.

Though our work is driven mainly by curiosity, its many applications have not escaped our attention. Lopsided dumbbell suspensions arise in the lipid encapsulation of mRNA into vaccines **(FIGURE 3** of [19]), meningococcal infections (**Fig. 1. B** of [20]), or when certain viral proteins interact with membranes (see **Figures 1.** through **3.** of [21]), or when synthesized deliberately from plastics [22,23]. Eq. (24) [or its dimensionless form Eq. (25)] are also main results of this Letter. Measurements of the complex viscosities of any of these systems could be used to determine $\lambda_H^\ell$ *versus* $\nu$. However, we know of no rheological characterizations of lopsided dumbbell suspensions of any kind.

Our work neglects interferences of the two Stokes flow velocity fields between the beads. Called *hydrodynamic interactions*, these interferences within lopsided dumbbells have yet to be analyzed, be the lopsided dumbbells rigid, or elastic. By contrast, these interferences are well-understood for symmetric dumbbells, both rigid (Subsection 14.6 of [2]; [24]) and elastic (Subsection 13.6 of [2]). We leave this intriguing problem for another day.

Hookean dumbbells are both infinitely extensible (as the force extending the dumbbell tends to positive infinity), and completely compressible (as the force compressing the dumbbell tends to negative infinity). By *completely compressible*, we mean that the lower bound for the center-to-center distance is zero. We leave the intriguing explorations of more exotic lopsided dumbbells, including both finitely extensible ones [such as FENE dumbbells (§ 13.5 of [2])], and incompletely compressible ones [such as Fraenkel dumbbells ([25,26,27]; Problem 13B.11 of [2])] for another day.

## II. ACKNOWLEDGMENT

This research was undertaken, in part, thanks to support from the Canada Research Chairs program of the Government of Canada for the Natural Sciences and Engineering Research Council of Canada (NSERC) Tier 1 Canada Research Chair in Physics of Fluids. This research was also undertaken, in part, thanks to support from the Discovery Grant program of the Natural Sciences and Engineering Research Council of Canada (NSERC) (A.J. Giacomin) and Vanier Canada Graduate Scholarship (M.A. Kanso).

Table 1: Nomenclature and Symbols. Legend: $M \equiv$ mass, $L \equiv$ length, and $t \equiv$ time.

|  | Unit | Symbol |
|---|---|---|
| Acceleration, $i$th bead | $L/t^2$ | $\ddot{\mathbf{r}}_i$ |
| Bead friction coefficient | $M/t$ | $\zeta$ |
| Bead friction coefficient, $i$th bead | $M/t$ | $\zeta_i$ |
| Bead mass | $M$ | $m$ |
| Bead radius | $L$ | $a$ |
| Bead radius, $i$th bead | $L$ | $a_i$ |
| Boltzmann constant | $ML^2/Tt^2$ | $k$ |
| Derivative of tensor $\mathbf{X}$, substantial | $\mathbf{X}/t$ | $\dfrac{d}{dt}\mathbf{X}$ |
| Derivative of tensor $\mathbf{X}$, upper convected | $\mathbf{X}/t$ | $\left\{\dfrac{d}{dt}\mathbf{X} - \mathbf{LX} - \mathbf{L}^T\mathbf{X}\right\}$ |
| Dirac delta function | $t^{-1}$ | $\delta(s)$ |
| Dumbbell center of resistance | $L$ | $\mathbf{R}^{(r)}$ |
| Dumbbell center-to-center vector | $L$ | $\mathbf{R}$ |
| Elastic shear modulus | $M/Lt^2$ | $nkT$ |
| Force on center of resistance, Brownian | $ML/t^2$ | $\mathbf{F}^{(b,cr)}(t)$ |
| Force on $i$th bead, random Brownian, | $ML/t^2$ | $\mathbf{F}_i^{(b)}(t)$ |
| Identity tensor | 1 | $\mathbf{I}$ |
| Kinetic molecular energy per molecule | $ML^2/t^2$ | $kT$ |
| Lopsidedness, dumbbell | 1 | $\nu \equiv a_2/a_1$ |
| Mass, $i$th bead, | $M$ | $m_i$ |
| Number density, dumbbells | $1/L^3$ | $n$ |
| Position vector, $i$th bead | $L$ | $\mathbf{r}_i$ |
| Real part of linear complex viscosity | $M/Lt$ | $\eta'(\omega)$ |

| Description | Units | Symbol |
|---|---|---|
| Relaxation time, lopsided Hookean dumbbell of center-to-center length $L$ | $t$ | $\lambda_H^\ell$ |
| Relaxation time, lopsided rigid dumbbell of center-to-center length $L$ | $t$ | $\lambda_R^\ell$ |
| Relaxation time, symmetric Hookean dumbbell of center-to-center length $L$ | $t$ | $\lambda_H$ |
| Relaxation time, symmetric rigid dumbbell of center-to-center length $L$ | $t$ | $\lambda_R$ |
| Separation of bead centers, rigid dumbbell | $L$ | $L$ |
| Solvent extra stress tensor | $M/Lt^2$ | $\tau_s$ |
| Spring stiffness, Hookean dumbbell | $M/t^2$ | $H$ |
| Stress tensor, dumbbell contribution to | $M/Lt^2$ | $\mathbf{S}^{(p)}$ |
| Temperature, absolute | $T$ | $T$ |
| Time | $t$ | $t$ |
| Time, interval of past | $t$ | $s \equiv t - t'$ |
| Velocity gradient gradient, at center of resistance | $L^{-1}t^{-1}$ | $\nabla\nabla\mathbf{u}^{(r)}$ |
| Velocity gradient, at center of resistance | $t^{-1}$ | $\nabla\mathbf{u}^{(r)}$ |
| Velocity gradient, at the center of resistance, transpose of | $t^{-1}$ | $\mathbf{L}$ |
| Velocity, bead $i$ | $L/t$ | $\dot{\mathbf{r}}_i$ |
| Velocity, fluid, at bead $i$ | $L/t$ | $\mathbf{u}_i$ |
| Velocity, higher order terms | $L/t$ | $HOT$ |
| Viscosity, dumbbell contribution to zero-shear | $M/Lt$ | $\eta_0 - \eta_s$ |
| Viscosity, solvent | $M/Lt$ | $\eta_s$ |
| Viscosity, zero-shear | $M/Lt$ | $\eta_0$ |

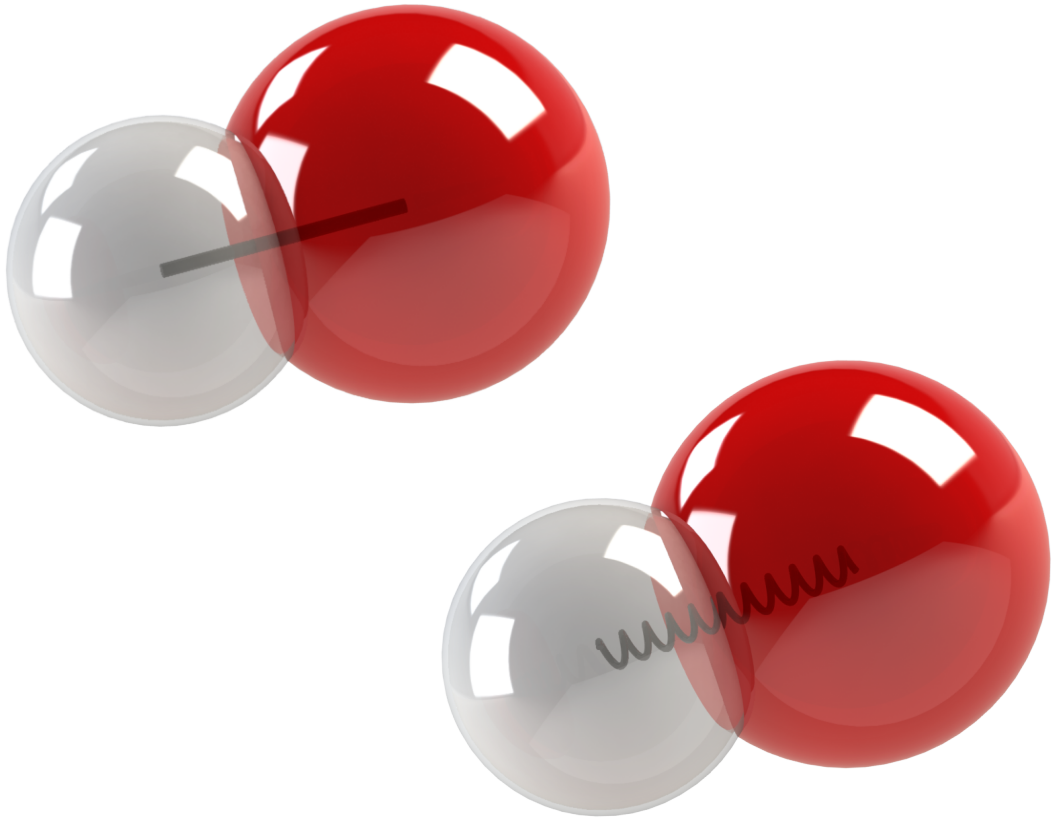

Figure 1: Rigid (upper left) and elastic (lower right) dumbbells both of lopsidedness $\nu = 5/4$, with bead interpenetration. Massless dimensionless rod (left) and Hookean spring (right) made visible, center-to-center, through translucent beads ($r_1/L = 1/3$, $r_2/L = 3/5$).

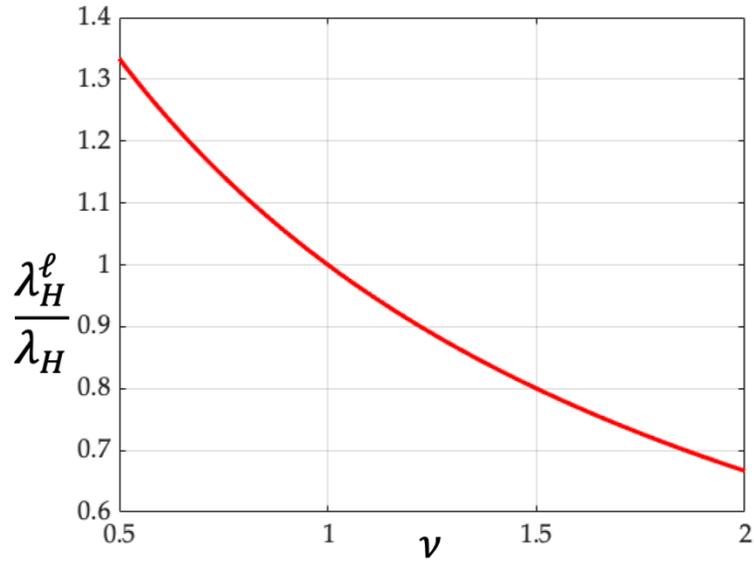

Figure 2: Descent of dimensionless relaxation time, $\lambda_H^\ell/\lambda_H$, with dumbbell lopsidedness, $\nu$ [Eq. (25) with Eq. (10)]. Where $\nu > 1$, the second bead is the larger of the two beads. Where $\nu < 1$, second bead is the smaller. ½<$\nu$ < 2.

## III. REFERENCES

[1] Bird, R.B., H.R. Warner Jr. and D.C. Evans, "Kinetic Theory and Rheology of Dumbbell Suspensions with Brownian Motion," *Adv. Polymer Sci.*, **8**, 1-90 (1971).

[2] Bird, R.B., C.F. Curtiss, R.C. Armstrong and O. Hassager, *Dynamics of Polymeric Liquids*, Vol. 2, 2nd ed., Wiley, New York (1987).

[3] Abdel-Khalik, S.I. and R.B. Bird, "Rheology of lopsided-dumbbell suspensions," *Applied Scientific Research*, **30**, 268-269 (1975).

[4] Bird, R.B., O. Hassager, R.C. Armstrong and C.F. Curtiss, *Dynamics of Polymeric Liquids*, Vol. 2, 1st ed., Wiley, New York (1977).

[5] Bird, R.B., A.J. Giacomin, A.M. Schmalzer and C. Aumnate, "Dilute Rigid Dumbbell Suspensions in Large-Amplitude Oscillatory Shear Flow: Shear Stress Response," *The Journal of Chemical Physics*, **140**(7), 074904 (2014), pp. 1-16. Errata: Ganged after Ref. 116 of [6] below.

[6] Saengow, C., A.J. Giacomin and C. Kolitawong, "Exact Analytical Solution for Large-Amplitude Oscillatory Shear Flow from Oldroyd 8-Constant Framework: Shear Stress," *Physics of Fluids*, **29**(4), 043101 (April, 2017), pp. 1-23. Errata: In column 4 of rows 12 and 13 of **TABLE IV.**, $\lambda_2 = \mu_0 = \mu_1$ should be $\mu_1 = -\lambda_1; \lambda_2 = \mu_0$ and $\mu_1 = \lambda_1; \lambda_2 = \mu_0$.

[7] Pak, M.C. (박명철), A.J. Giacomin and M.A. Kanso (منى قانصو), "Steady Elongational Flow from Rotarance Theory," *Physics of Fluids*, **35**(10), Part 2 of 4, 103116 (October, 2023), pp. 1-15. Feature article.

[8] Phan-Thien, N. and N. Mai-Duy, *Understanding Viscoelasticity, Basics of Rheology*, 1st ed., Springer, New York (2002).

[9] Phan-Thien, N., *Understanding Viscoelasticity, An Introduction to Rheology*, 1st ed., Springer, New York (2013).

[10] Phan-Thien, N. and N. Mai-Duy, *Understanding Viscoelasticity, An Introduction to Rheology*, 3rd ed., Springer, New York (2017).

[11] Lemons, D.S. and A. Gythiel, "Paul Langevin's 1908 paper ''On the Theory of Brownian Motion'' [Sur la théorie du mouvement brownien]," *American Journal of Physics*, **65**(11), 1079–1081 (1997).

[12] Langevin, P., "Sur la théorie du mouvement brownien," *Comptes rendus de l'Académie des Sciences*, **146**, 530-533 (1908).

[13] Lumley, J.L., "Applicability of the Oldroyd constitutive equation to flow of dilute polymer solutions," *The Physics of Fluids* **14**(11) 2282-2284 (1971).

[14] Lumley, J.L., "Erratum: Applicability of the Oldroyd constitutive equation to flow of dilute polymer solutions," *The Physics of Fluids*, **15**(11), 2081-2081 (1972).

[15] Tanner, R.I., "*Engineering Rheology,*" revised ed., Clarendon Press, Oxford (1988).

[16] Bird, R.B., R.C. Armstrong and O. Hassager, *Dynamics of Polymeric Liquids*, Vol. 1, 1st ed., Wiley, New York (1977).

[17] Saengow, C. and A.J. Giacomin, "Ongoing Relevance of Oldroyd 8-Constant Fluids," *Journal of Non-Newtonian Fluid Mechanics*, **299**, 104653 (2022), pp. 1-9.

[18] Bird, R.B., O. Hassager, R.C. Armstrong and C.F. Curtiss, *Dynamics of Polymeric Liquids*, Vol. 2, 1st ed., Wiley, New York (1977).

[19] Brader, M.L., S.J. Williams, J.M. Banks, W.H. Hui, Z.H. Zhou and L. Jin, "Encapsulation state of messenger RNA inside lipid nanoparticles," *Biophysical Journal*, **120**, 2766–2770 (July 20, 2021).

[20] Manno, D.E., A. Talà, M. Calcagnile, S.C. Resta, P. Alifano and A. Serra, "Surface architecture of Neisseria meningitidis capsule and outer membrane as revealed by atomic force microscopy," *Research in Microbiology*, **172**(6), 103865 (2021).

[21] Alimohamadi, H., E. W.-C. Luo, S. Gupta, J. de Anda, R. Yang, R. Mandal and G.C.L. Wong, "Comparing multifunctional viral and eukaryotic proteins for generating scission necks in membranes," doi: https://doi.org/10.1101/ 2024.01.05.574447, *bioRxiv*, 2024.01.05.574447 (January 7, 2024).

[22] Peng, B., H.R. Vutukuri, A. van Blaaderen and A. Imhof, "Synthesis of fluorescent monodisperse non-spherical dumbbell-like model colloids," *Journal of Materials Chemistry*, **22** (41), 21893-21900 (2012).

[23] Kim, J.-W., R.J. Larsen and D A. Weitz, "Synthesis of nonspherical colloidal particles with anisotropic properties," *Journal of the American Chemical Society*, **128**(44), 14374-14377 (2006).

[24] Phan-Thien, N. and R.I. Tanner, "Langevin analysis of dumbbells with hydrodynamic interaction," *Rheologica Acta*, **17**, 568-577 (1978).

[25] Fraenkel. G.K., "Visco-elastic effect in solutions of simple particles," *J. Chem. Phys.*, **20**, 642-647 (1952). Errata: In Eq. (21), "$12\tau_0$" should be "$12\omega\tau_0$"; Eq. (17a) should be $s = (1/\epsilon)\left[(2r/a) - 1\right]$.

[26] Fraenkel, G.K., "The viscosity and shear elasticity of solutions of simple deformable particles," PhD Thesis, Cornell University, Cornell, NY (1949).

[27] Piette, J.H., C. Saengow and A.J. Giacomin, "Zero-Shear Viscosity of Fraenkel Dumbbell Suspensions," *Physics of Fluids*, **32**(6), 063103 (June, 2020), pp. 1-11.